\documentclass[twocolumn, prl, aps, floatfix, superscriptaddress, longbibliography]{revtex4-1}

\usepackage{color}
\usepackage{graphicx}
\usepackage{physics}
\usepackage{amsthm}
\usepackage{amsmath}
\usepackage{amssymb}
\usepackage{enumerate}
\usepackage{placeins}
\usepackage{booktabs}
\usepackage{dsfont}
\usepackage{yfonts}
\usepackage{hyperref}
\usepackage{bbold}
\usepackage{upgreek}
\usepackage{hhline}
\usepackage{listings}
\usepackage{siunitx}

\usepackage[dvipsnames]{xcolor}

\AtBeginDocument{%
	\heavyrulewidth=.08em
	\lightrulewidth=.05em
	\cmidrulewidth=.03em
	\belowrulesep=.65ex
	\belowbottomsep=0pt
	\aboverulesep=.4ex
	\abovetopsep=0pt
	\cmidrulesep=\doublerulesep
	\cmidrulekern=.5em
	\defaultaddspace=.5em
}

\newif\ifarXiv
\arXivtrue 

\ifarXiv

\usepackage{pdfpages} 
\usepackage{pgffor} 
\usepackage{xr} 

\makeatletter
\AtBeginDocument{\let\LS@rot\@undefined}
\makeatother

\def\supplementfilename{MIT_supplement}

\pdfximage{\supplementfilename.pdf}
\def\numbersupplementpages{\the\pdflastximagepages}

\fi 

\begin{document}
	
\title{Metal-insulator transition in a boundary three chain model}
	
\author{Niels John} 
\affiliation{Institut f\"ur Theoretische Physik, Universit\"at Leipzig, D-04103, Leipzig, Germany} 
	
\author{Yuval Gefen}
\affiliation{Department of Condensed Matter Physics, Weizmann Institute, 7610001 Rehovot, Israel}

\author{Bernd Rosenow}
\affiliation{Institut f\"ur Theoretische Physik, Universit\"at Leipzig, D-04103, Leipzig, Germany} 
	
\begin{abstract}

We study the boundary physics of bulk insulators by considering three coupled Hubbard chains in a linear confining potential. In the Hartree-Fock approximation, the ground state at and slightly off the particle-hole symmetric point 
 remains insulating even at large slopes of the confining potential.  By contrast, accounting for quantum fluctuations and correlations through  a combination of bosonization and RG methods, we find that there is always a gapless dipole mode, but it does not contribute to a finite compressibility. Moreover, when increasing the slope of the confining potential at half filling, we find a quantum phase transition between an insulating and a metallic state, indicating the formation of a soft edge state of non-topological origin. Away from half filling, a similar transition takes place. 	
	
\end{abstract}
	
\maketitle


\noindent \textit{Introduction:} The boundaries of two-dimensional (2D) phases of matter may play a pivotal role when the bulk spectrum is gapped \cite{Halperin1982, Laughlin1981, Kane2015, Bernevig2006, Hasan2010, Qi2011, Koenig2007}. Having a gapless spectrum at the edge may render the system conducting underpinned by bulk topological invariants \cite{Altland1997, Wen2007}. Examples range from quantum Hall phases (integer, fractional, non-Abelian) featuring chiral modes \cite{Halperin1982, Laughlin1981}, helical modes \cite{Kane2005, Bernevig2006}, and other 2D topological phases \cite{Kosterlitz1972, Kosterlitz1973, Haldane1988}. On the other hand, interaction driven Mott insulators  \cite{Mott1949, Schulz} are usually not characterized by conducting edge modes.  

A possible path towards obtaining a clearer insight into 2D boundary physics would be to study interacting models comprising few-chains positioned in a potential that mimicks the confinement at the edge. 
Within the HF approximation, for a system of several coupled chains an alternation of compressible and incompressible regions has been found \cite{Khanna.2021}. In the absence of a confining potential, bosonization was employed for both single and multiple chains \cite{Schulz, Balents1996, Hur2001, GiamarchiBook, FinkelStein1993, Miao2016,Tsuchiizu2001}.
For a single Hubbard chain there exists an insulating phase without magnetic order. Considering several coupled Hubbard chains one finds a variety of instabilities as a function of interchain hopping, as well as Hubbard and next-nearest-neighbor interaction strength \cite{Schulz, Balents1996, Hur2001, GiamarchiBook, FinkelStein1993, Miao2016}. For two coupled chains the metal-insulator transition was observed as functions the chemical potential \cite{Tsuchiizu2001}, the filling \cite{Balents1996}, the interchain hopping \cite{Balents1996, Hur2001, Tsuchiizu2001}, or the strength of on-site and nearest neighbor interactions \cite{Tsuchiizu2001, Hur2001}. Superconducting $d$-wave or $s$-wave states or charge modulated states have been observed within the framework of two coupled chains as well \cite{GiamarchiBook}.

\begin{figure}[t!]
	\centering
	\includegraphics[width=0.45\textwidth]{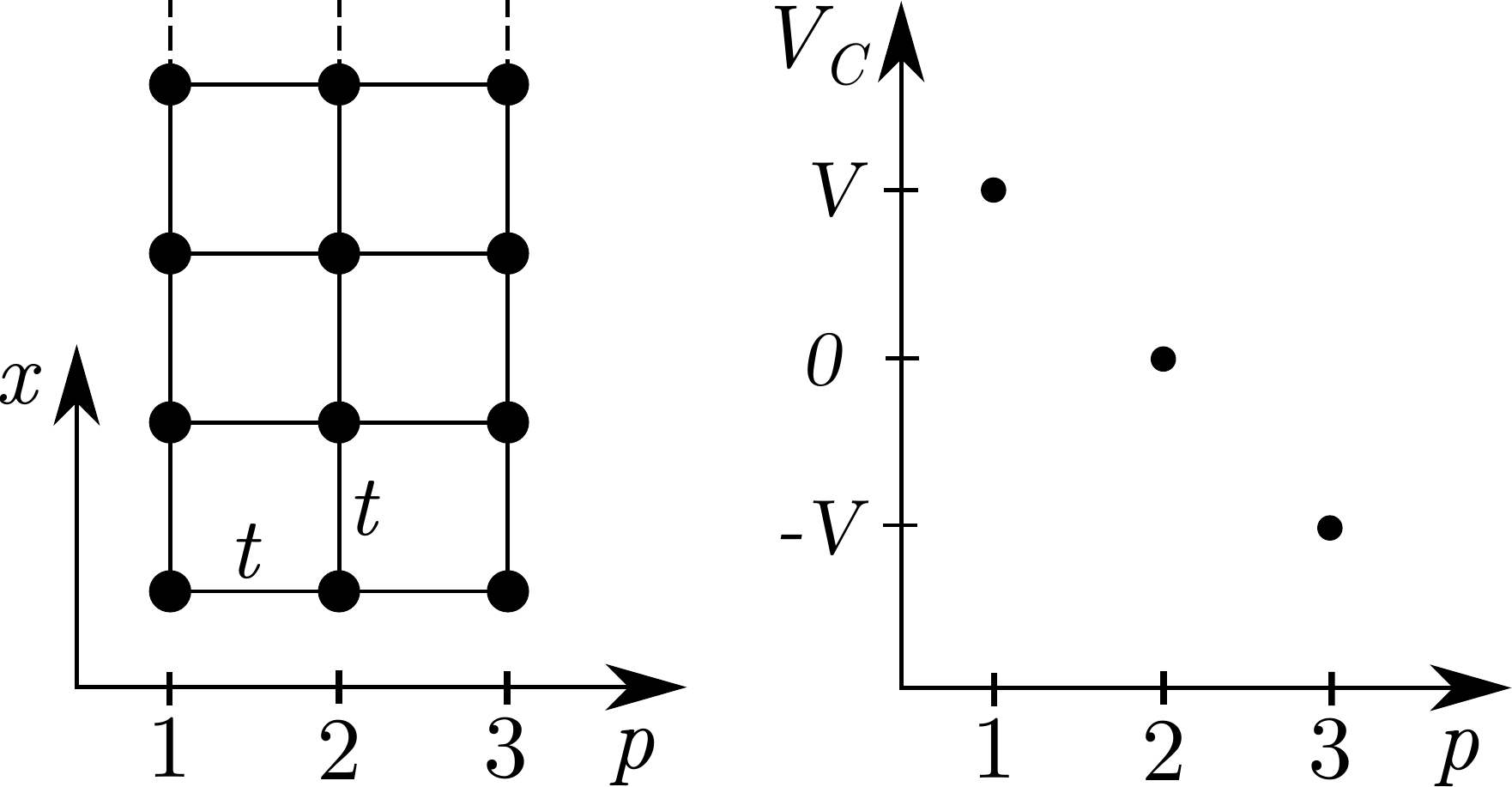}
	\caption{Geometry of three coupled Hubbard chains subject to a confining potential. In the left panel, the lattice structure with the single-electron hoppings $t$ across and along the chains is depicted. The right panel shows the confining potential as a function of the chain index $p$.}
	\label{fig:Fig1}
\end{figure}

\begin{figure}[h!]
	\centering
	\includegraphics[width=0.45\textwidth]{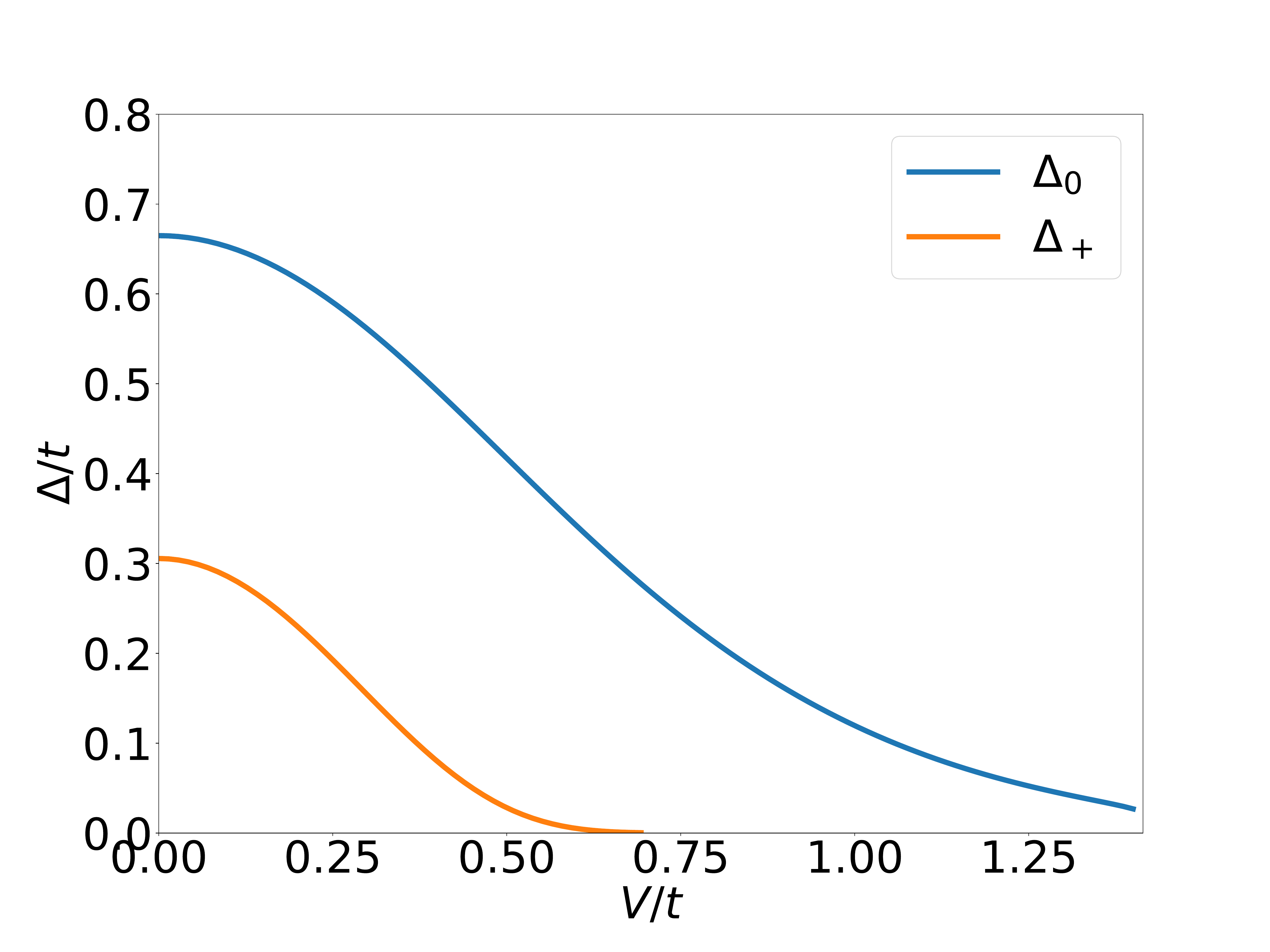}
	\caption{Gaps of the two   modes contributing to charge density and current as a function of the the confining potential strength $V$ at half filling. $\Delta_{+}$ and $\Delta_{0}$ are the charge gaps of the fields $\phi_{\rho,+}$ and $\phi_{\rho,0}$. The latter remains finite up until $V/t=\sqrt{2}$, whereas the former drops to zero at $V/t \simeq 0.7$. This marks the transition from an incompressible state into a compressible state which is expected to have a finite conductance. The transition happens at the particle-hole symmetric point and persists for a finite range of values of the chemical potential.}
	\label{fig:Fig2}
\end{figure}

Here we take up the challenge of modelling the boundary of interacting or strongly correlated two-dimensional materials. Specifically, we consider a model of three coupled Hubbard chains under the influence of an external confining potential, mimicking the edge of a 2D Mott insulator with smooth confinement. In addition to a standard nearest-neighbor Hamiltonian with Hubbard interactions, we assume a chain dependent potential which mimics a confining potential in the perpendicular $y$-direction. The system is depicted in \autoref{fig:Fig1}. Employing bosonization technique tools, we study the collective phases of this many fermion system. Accounting for the interplay of the interaction strength and the confining potential, we find a novel mechanism of a metal-insulator transition. This transition occurs as function of the confining potential, not withstanding the fact that the excitation spectrum remains gapless. Our results are depicted in \autoref{fig:Fig2}. Note that despite the presence of a gapless mode, the system can be an insulator since that mode is an interchain plasmon not contributing to the compressibility and conductance.
The overall filling of the system is determined by the chemical potential $\mu$. As a first step towards a realistic model for the edge region of a two-dimensional Mott insulator, we allow for a confining potential $V_C(p)$. The confining potential is different for chains with different index $p$ and is constant along each individual chain. In the following, we will be mostly concerned with the case that the confining potential is split symmetrically around the chemical potential, meaning that $V_{\text{C}}(1)=-V_{\text{C}}(3) \equiv V$ and $V_{\text{C}}(2) = 0$. The full Hamiltonian of the system is 
\begin{align} \label{eq:CoupledChainsHamiltonian}
	\begin{split}
		H &= -t\sum_{j,\sigma,p} \left(c^{\dagger}_{j,p,\sigma}c^{\phantom{\dagger}}_{j+1,p,\sigma} + \text{h.c.}\right) - \mu \sum_{j,\sigma,p} n_{j,\sigma,p}\\
		&\ \ \ - t \sum_{j,\sigma} \sum_{p=1,2} \left(c^{\dagger}_{j,p,\sigma}c^{\phantom{\dagger}}_{j,p+1,\sigma} + \text{h.c.}\right) \\
		&\ \ \ + V\sum_{j,\sigma} \left(c^{\dagger}_{j,1,\sigma} c^{\phantom{\dagger}}_{j,1,\sigma} - c^{\dagger}_{j,3,\sigma} c^{\phantom{\dagger}}_{j,3,\sigma}\right) \\
		&\ \ \ + U\sum_{j,p} n_{j,p,\uparrow}n_{j,p,\downarrow} \ ,
	\end{split}
\end{align}
where $t$ is the nearest-neighbor hopping strength. Due to the tunneling between the chains, the chain index $p$ is not  a good quantum number.

For the non-interacting system with $U=0$, diagonalization of $H$ yields a Hamiltonian with three bands and pseudospin indices $\tau = -1,0,+1$. The dispersions are $E_{k,\tau} = -2t\cos(ak) + \tau\sqrt{2t^2 + V^2}$, such that  at half-filling and weak confining potential $V/t\ll 1$, the Fermi momenta of these bands are $k_{F,\tau} = \pi/2 -\tau(\pi/4+V^2/t^2)$. Interestingly, one band remains at exactly half-filling, independent of the actual value of $V$. Once the slope of the confining potential reached the critical value $V/t=\sqrt{2}$, the lower band with $\tau = -1$ is completely filled and the upper band with $\tau=+1$ is empty. Thus, for values $V/t>\sqrt{2}$ and without interaction, the system is trivial in the sense that only the middle chain remains at half-filling, whereas the one with $p=3$ is completely filled and thus similar to a bulk state, and the outermost chain with $p=1$ is empty.

\noindent \textit{Hartree-Fock analysis:} In order to gain some intuition into the interacting theory, we use Hartree-Fock (HF) analysis to determine the band structure self-consistently. We reexpress the interaction term $n_{j,p,\uparrow} n_{j,p,\downarrow}$ as $ \langle  n_{j,p,\uparrow} \rangle_0  n_{j,p,\downarrow} +   n_{j,p,\uparrow}  \langle n_{j,p,\downarrow}\rangle_0 -  \langle n_{j,p,\uparrow}\rangle_0  \langle n_{j,p,\downarrow}\rangle_0$, where $\langle\dots \rangle_0$ denotes the expectation value with respect to the HF ground state. We perform a restricted HF analysis with
\begin{align}
	\langle n_{j,p,\sigma} \rangle_0  = \frac{n_p}{2} + \sigma S_p(-1)^j \ ,
\end{align}
where $n_p$ is the average density in chain $p$ and the order parameter $S_p$ is the staggered magnetization of chain $p$. The effective single-particle Hamiltonian in terms of the order parameter is discussed in the supplement \cite{supp}. The densities and staggered magnetizations are determined selfconsistently. 

For a single half-filled Hubbard chain, the antiferromagnetic order parameter $S$ is finite for any $U>0$ and the system develops a finite band gap. As the chain is doped away from half filling, the order parameter drops to zero rendering the system compressible. Then, the band gap closes and the system has two Fermi points at $k = \pm k_F$. We can use a similar picture for the analysis of the three coupled chains: each pair of Fermi points corresponds to one gapless charge mode. Within HF theory and near half filling, the gapping of Fermi points is only possible due to a finite antiferromagnetic magnetization.

\begin{figure}[t!]
	\centering
	\includegraphics[width=\linewidth]{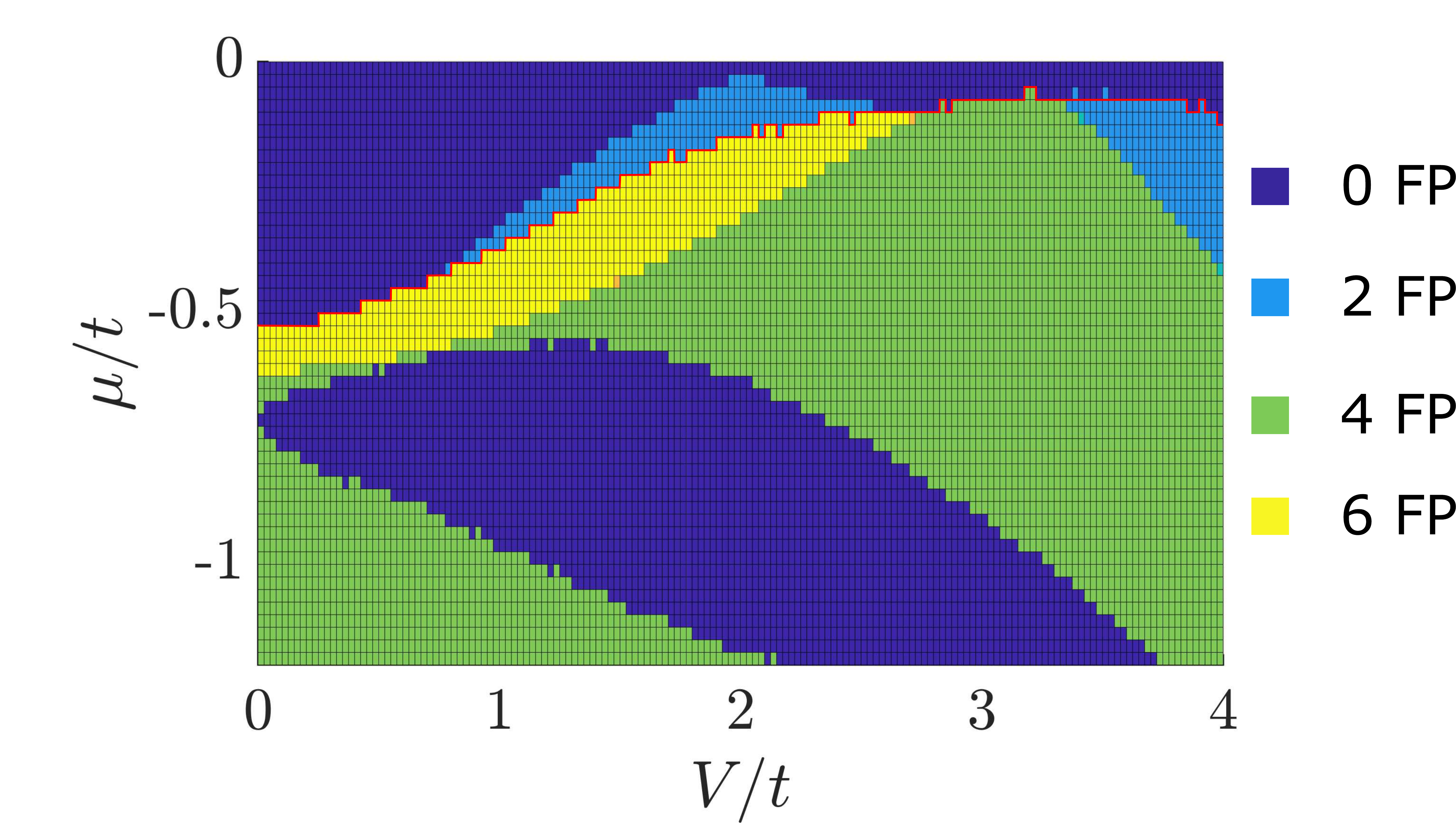}
	\caption{Mean field phase diagram for $U/t=5$. The plot shows the number of Fermi points of the self-consistently determined band structure as a function of the chemical potential $\mu/t$ and the confining strength $V/t$. The red line indicates the transition between anti-ferromagnetic order and a paragmagnet. As an interpretation, we can identify a pair of Fermi points with an ungapped (charge) mode.}
	\label{fig:Fig3}
\end{figure}

In \autoref{fig:Fig3}, the number of Fermi points as a function of the chemical potential and the confining potential is shown. Interestingly, the half-filled system remains gapped and has no Fermi points, independent of the  value of $V$. This constitutes an adiabatic connection from the Mott insulating state at $V=0$ to an asymptotic state $V/t \gg 1$ where one chain is completely empty, one chain is completely filled and the middle chain is still a gapped Mott insulator.

We observe transitions into states with a finite number of Fermi points as a function of $\mu$ and $V$. At a finite $\mu/t \simeq -0.2$, increasing $V$ eventually leads to a transition into a state with two Fermi points, which is the equivalent of one gapless charge mode. At the transition point, the band gap closes and the system becomes compressible. If we increase $V$ even further, the system transitions from a state with two Fermi points into one with six Fermi points. In the latter state, not only the band gap vanishes, but the antiferrogmagnetic magnetization vanishes as well. The regime with six Fermi points, or three ungapped charge modes, is thus equivalent to a system of non-interacting electrons.

Depending on the actual value of $V$, we observe similar transitions as $\mu$ varies. At  $V=0$, lowering $\mu$ leads to a transition into a gapless state with three Fermi points around $\mu/t \simeq -0.55$. As before, in this gapless regime  the antiferrogmagnetic order parameter vanishes. If, however $1<V/t<2$, a variation of $\mu$ leads to similar transitions as described in the paragraph above: first, the system transites into a state with two Fermi points which is accompanied by a band gap closing. For even lower $\mu$, we observe the same transition into a compressible, non-interacting state with six Fermi points and vanishing antiferromagnetic order parameter. 

We emphasize that the absence of a gap closing at half filling is an important difference between three coupled and three uncoupled chains placed in a linear confinging potential. In the former case, all three chains always have a finite band gap. In the latter case and for sufficiently large $V$, the chemical potential with respect to the bottom of the band for a given chain will exceed the band gap and the respective order parameters drop to zero. This will render the aforementioned chains compressible and, hence, make the whole system compressible despite the fact that it remains at exactly half filling.

\noindent \textit{Bosonization:} We use the bosonization technique and a first-order renormalization group analysis to investigate the interacting system with $U>0$. For $V/t<\sqrt{2}$, the three bands are bosonized individually, such that the kinetic energy is diagonal in the $\tau$-basis. For each band, there are four bosons fields $\phi_{\nu,\tau}$ and $\theta_{\nu,\tau}$ with commutation relations  $[\phi_{\nu,\tau}(x),\partial_{x'}\theta_{\nu',\tau'}(x')] = i \,\hbar \, \pi\, \delta_{\nu,\nu'}\delta_{\tau,\tau'}\delta(x-x')$, where $\nu = \rho, \sigma$ labels the charge or spin channel. As usual, the density of band $\tau$ is related to the fields as $\rho_{\tau}(x) = \rho_{\tau,0} + (\sqrt{2}/\pi) \cdot \partial_x\phi_{\rho,\tau}(x)$, where $\rho_{0,\tau}$ is the average density of the band. The relation between the densities of the bands and the densities of the chains is shown in the supplement \cite{supp}.
Keeping only the most relevant contributions of the interaction terms, we can write it as $H_{\text{int}} = H_{\text{int}}^0 + H_{\text{int}}^{2k_F}$, where the first term is a density-density interaction in the new basis and the second term describes Umklapp scattering.  We note that in  $H_{\text{int}}^0$ all three fields are coupled among each other. 

To partially decouple $H_{\text{int}}^0$ we introduce the symmetric and antisymmetric combination of the fields $\phi_{\nu, \tau=+1}$ and $\phi_{\nu, \tau=-1}$. The fields are transformed as
\begin{align} \label{eq:TrafoO2}
	\left(\begin{array}{c}
		\phi_{\nu,\alpha=+} \\
		\phi_{\nu,\alpha=0}  \\
		\phi_{\nu,\alpha=-} 
	\end{array}\right)
	= \left(
		\begin{array}{ccc}
			\frac{1}{\sqrt{2}} & 0 & \frac{1}{\sqrt{2}} \\
			0 & 1 & 0 \\
			\frac{1}{\sqrt{2}} & 0 & -\frac{1}{\sqrt{2}}
		\end{array}
		\right)
	\left(\begin{array}{c}
		\phi_{\nu,\tau =+1} \\
		\phi_{\nu,\tau =0} \\
		\phi_{\nu,\tau =-1}
	\end{array}\right) \ .
\end{align}
The $\theta$ fields are transformed accordingly. The new field index takes values $\alpha=+,0,-$. In the $\alpha$ basis, the Hamiltonian $\mathcal{H}_0 = H_0 + H^0_{\text{int}}$ is given by
\begin{align} \label{eq:BosonizedHamiltonianI}
	\begin{split}
		\mathcal{H}_0 &=  \frac{1}{2\pi}\sum_{\alpha} \sum_{\nu = \rho,\sigma}  \left[ u_{\nu,\alpha}K_{\nu,\alpha}\left(\partial_x \theta_{\nu, \alpha}\right)^2 + \frac{u_{\nu,\alpha}}{K_{\nu,\alpha}} \left(\partial_x \phi_{\nu, \alpha}\right)^2\right] \\
		&+ \frac{\gamma}{\pi} \sum_{\nu = \rho,\sigma}   \left(\partial_x \phi_{\nu,+} \right)  \left(\partial_x \phi_{\nu,0} \right)   \ .
	\end{split}
\end{align}
with $\gamma = \left( U /(t) \cdot \sqrt{2}\left(2 (V/t)^2+1\right)\right)/\left((V/t)^2+2\right)^2$ and 
\begin{align}
	u_{\nu}K_{\nu}\ \hat{=}  \left(
		\begin{array}{ccc}
			\sqrt{2-(V/t)^2} & 0 & 0 \\
			0 & 2 & 0 \\
			0 & 0 & \sqrt{2-(V/t)^2}
		\end{array}
		\right) \ .
\end{align}
We can see from Eq. \eqref{eq:BosonizedHamiltonianI} that the transformation from Eq.~{\eqref{eq:TrafoO2}} only decouples the field $\phi_{\nu,-}$ from the other two in the kinetic Hamiltonian. In the new basis, the two  most relevant Umklapp terms are
\begin{align} \label{eq:UmklappTerms}
	\begin{split}
		H^{2k_F}_{\text{int}} &=  -\frac{g_3^{(0,0;0,0)}}{2\pi^2a^2} \cos(2\sqrt{2}\phi_{\rho,0} - \delta_0\cdot  x)   \\ 
		&\ \ \ -  \frac{g_3^{(1,-1;-1,1)}}{2\pi^2a^2}  \cos(2\theta_{\sigma,-}) \cos(2 \phi_{\rho,+} -\delta_+\cdot x) \ ,
	\end{split}
\end{align}
where $\delta_+ = \delta_0 \equiv 0$ at half filling. The remaining relevant Umklapp terms contain linear combinations of $\phi_{\rho,0}$ and $\phi_{\rho,+}$ as arguments and do not contribute additional physics for the charge modes. The bare values of the coupling constants can be found in the supplemental material \cite{supp}. 

In order to derive the Umklapp terms above, we use refermionization \cite{GiamarchiBook, vonDelft1998}, the fermionic field operators can be expressed  as exponentials of the boson fields in the $\tau$-basis, multiplied by a  plane wave with Fermi momentum $k_{F,\tau}$:
\begin{align}\label{eq:FieldOperator}
	\psi_{r,\tau,s}(x) \propto e^{irk_{F,\tau}x}e^{-i/\sqrt{2}(r\phi_{\rho,\tau} - \theta_{\rho,\tau}+s(r\phi_{\sigma,\tau}-\theta_{\sigma,\tau}))} \ ,
\end{align}
where $r=\pm1$ for right- and left-movers and $s = \pm1$ for spin up and spin down. To obtain the  Umklapp terms given in Eq.~\eqref{eq:UmklappTerms}, we rewrite the interaction term in Eq.~\eqref{eq:CoupledChainsHamiltonian} in terms of  bosonic exponentials, leaving us with terms of the type $\cos(c_{\beta}\cdot \phi_{\rho,\beta} + \delta_{\beta} \cdot x)$, where $c_{\beta}$ is a real number and the displacement $\delta_{\beta}$ is the sum or the difference of all the plane wave momenta. For the Umklapp terms from Eq.~\eqref{eq:UmklappTerms}, the displacement is $\delta_{\beta} = \pm 2\pi$ at half filling. By construction, the boson fields vary slowly on the scale of inverse Fermi momentum. Indeed, if $\delta_{\beta} \neq 0$ modulo $2\pi$ the cosine operators oscillate on length scales much shorter than $1/k_{F,\tau}$ and thus vanish upon integration over $x$. This restricts the processes that we should include into the bosonization framework, and, in particular, leaves us with only a few Umklapp terms we need to take into consideration. 

Remarkably, the charge field $\phi_{\rho,-}$ does not appear in the Umklapp terms. This is a direct consequence of the fact that  only cosines with vanishing displacements should be kept in the Hamiltonian. Utilizing Eq.~$\eqref{eq:FieldOperator}$, we note that cosines with arguments proportional to $\phi_{\rho,-} $ \textit{always} come with a displacement $\delta_{-} = \pm2k_{F,+1} \mp 2k_{F,-1} \equiv \pm \pi$. Keeping in mind that the position variable is understood as $x = j\cdot a$ with $j\in \mathbb{N}$, these cosines change sign on adjacent lattice sites and vanish upon integration over $x$. 

Bosonic fields appearing as the argument of an RG relevant cosine term are locked in one of the extrema of the cosine, giving the respective boson field a finite expectation value. Expanding the cosine about this minimum, a mass term quadratic in the boson field is obtained, implying that the boson field develops a gap. According to the discussion above, $\phi_{\rho,-}$ remains gapless such that one might conclude that the coupled chains remain compressible. However, $\phi_{\rho,-}$ does not appear in the charge density
\begin{align} \label{eq:ChargeDensity}
	\rho(x) = \rho_0 +\frac{\sqrt{2}}{\pi} \left(\partial_x\phi_{\rho,0} \right) + \frac{2}{\pi} \left(\partial_x\phi_{\rho,+} \right)  \ ,
\end{align}
where $\rho_0 = 1$ at half filling. The absence of $\phi_{\rho,-}$ in the charge density implies that the conjugate field will not show up in the associated current density. Invoking this argument, the system can still be an insulator as long as the fields $\phi_{\rho,+}$ and $\phi_{\rho,0}$ are gapped. 

To decouple the remaining two fields, we first rescale the fields $\phi_{\nu,0}\mapsto \left(2/\sqrt{2-V^2}\right)^{-1/2} \phi_{\nu,0}$ and $\theta_{\nu,0}\mapsto \left(2/\sqrt{2-V^2}\right)^{1/2} \theta_{\nu,0}$ and diagonalize the remaining Hamiltonian. The rescaling is necessary in order to obey the bosonic commutation relations for the new linear combinations $\tilde{\phi}_{\nu,\alpha}$ of the fields. In matrix notation, we express the transformation as
\begin{align} \label{eq:FieldTrafoO3}
	\left(\begin{array}{c}
		\phi_{\nu,+} \\
		\phi_{\nu,0}  \\
		\phi_{\nu,-} 
	\end{array}\right)
	= \left(
		\begin{array}{ccc}
			\eta_{\nu;++} & \eta_{\nu;0+} & 0 \\
			\eta_{\nu;+0} & \eta_{\nu;00} & 0 \\
			0 & 0 & 1
		\end{array}
		\right)
	\left(\begin{array}{c}
		\tilde{\phi}_{\nu,+} \\
		\tilde{\phi}_{\nu,0} \\
		\tilde{\phi}_{\nu,-}
	\end{array}\right) \ .
\end{align}
After diagonalization, the full Hamiltonian including the relevant sine-Gordon terms is $H  = \tilde{\mathcal{H}}_0 + \tilde{\mathcal{H}}_{\text{int}}^{2k_F}$ with
\begin{align} \label{eq:FinalDiagonalHamiltonian}
	\tilde{\mathcal{H}}_0 = \frac{1}{2\pi}\sum_{\alpha,\nu = \rho,\sigma}  \left[ \tilde{u}_{\nu,\alpha}\tilde{K}_{\nu,\alpha}\left(\partial_x \tilde{\theta}_{\nu, \alpha}\right)^2 + \frac{\tilde{u}_{\nu,\alpha}}{\tilde{K}_{\nu,\alpha}} \left(\partial_x \tilde{\phi}_{\nu, \alpha}\right)^2\right] \ ,
\end{align}
and $\tilde{\mathcal{H}}^{2k_F}_{\text{int}}$ is the same as Eq.~\eqref{eq:UmklappTerms} except the fields are transformed according to Eq.~\eqref{eq:FieldTrafoO3}. Further, $\tilde{u}_{\nu,\alpha}\tilde{K}_{\nu,\alpha} = \sqrt{2-(V/t)^2}$. For $\tilde{u}_{\nu,\alpha}/\tilde{K}_{\nu,\alpha}$ see the supplemental material \cite{supp}. 

Without confining potential $V/t=0$, both of the sine-Gordon terms in Eq.~\eqref{eq:UmklappTerms} are relevant, meaning that the fields $\phi_{\rho,+}$ and $\phi_{\rho,0}$, and $\theta_{\sigma,-}$ are gapped. In the following, we study the charge gaps for the fields $\phi_{\rho,+}$,  $\phi_{\rho,0}$ as a function of the confining potential strength $V/t$. 
We calculate the gaps gaps utlizing a variational approach outlined in \cite{GiamarchiBook}. The gap for each charge mode is given by
\begin{align} \label{eq:Gap}
	\Delta^{(\gamma)} =  u^{(\gamma)}_{\rho}\Lambda \left(\frac{4K^{(\gamma)}_{\rho} y^{(\gamma)}}{(\alpha\Lambda)^2}\right)^{1/(2-2K^{(\gamma)}_{\rho})} \ , 
\end{align}
where $K^{(\gamma)}_{\rho}$ is obtained from the scaling dimension of the respective sine-Gordon term, $u^{(\gamma)}_{\rho}$ is the excitation velocity, and $\gamma$ is a multi-index that takes values $\gamma = 0,0;0,0$ and $\gamma = 1,-1;-1,1$. In the following discussion, we use the abbreviations $\Delta_0 \equiv \Delta^{(0,0;0,0)}$ and $\Delta_+ \equiv \Delta^{(1,-1;-1,1)}$, such that $\Delta_0$ is the charge gap for the field $\phi_{\rho,0}$, and $\Delta_+$ for 
the field $\phi_{\rho,+}$ .  $\Lambda=\pi/(2a)$ is a cutoff in momentum space,  and $y^{(\gamma)} = g_3^{(\gamma)}/(\pi\hbar v_F)$ is the dimensionless coupling constant. We note that the renormalization of the bare gaps  inside the brackets in Eq.~\eqref{eq:Gap} happens due to the exponent $1/(2-2K^{(\gamma)}_{\rho})$, which is the inverse of the scaling dimension of the corresponding term. Since both $u_{\rho}$ and $K_{\rho}$ depend on $V/t$, we 

The magnitudes of the gaps as a function of $V/t$ are shown in \autoref{fig:Fig2}. One sees that both charge gaps are finite but different in magnitude at $V/t=0$. Further, we find that  the magnitudes of the gaps calculated in Hartree-Fock theory agree quite well with the result from bosonization. As $V/t$ increases, both gaps decrease in magnitude. Again, this is in agreement with the results obtained from HF. At $V/t \simeq 0.7$, the charge gap $\Delta_+$ drops to zero. The closing of this gap is accompanied by the corresponding sine-Gordon term from Eq.~\eqref{eq:UmklappTerms} becoming an irrelevant operator. More importantly, we observe a metal-insulator transition as a function of the confining potential strength. This is an important difference compared to HF. For the latter, at the symmetry point $\mu = 0$ the system remains gapped independent of the actual value of $V/t$. The charge gap $\Delta_0$ remains finite all the way up to $V/t\simeq \sqrt{2}$.

\noindent \textit{Discussion:} 
We have considered a simplified picture of an edge of a strongly correlated 2D electron gas, modelled by three coupled chains subject to a “confining potential”. Our bosonization approach is compared with a HF analysis. For the latter, only when being away from the particle-hole symmetry point, we find that as we increase the steepness of the confining potential, we transition from a gapped phase to an ungapped one, from a vanishing to a finite conductance phase, and from a finite to a  zero magnetization phase. The magnetization transition does not coincide with the other two. By contrast, our bosonization analysis, both at and away from the symmetry point, reveals distinctly different features.  As we increase the steepness of the confining potential, we observe a transition from a non-conducting to a conducting phase (the spectrum remains  gapless throughout, and on both sides of the transition the phases are non-magnetic.)

This more reliable bosonization analysis combined with renormalization group insights, which accounts for quantum fluctuations, reveals that the phase transition reported here takes place notwithstanding the fact that the spectrum is always gapless. The insulating-to-conducting phase transition is accompanied by a gap closing of a collective charge mode as the confinement strength increases.
We thus find that as a function of the confining potential strength, there can be a quantum phase transition at the boundary of interaction driven Mott insulators, with experimentally observable consequences. 

Our prediction of a gap closing in an interacting system as a function of the slope of a confining potential can  be tested with present day experimental techniques.  In bilayer graphene, applying a perpendicular electric field can open up a bulk gap at the neutrality point  \cite{Li.2018}.  A combination of 
 top gates and a back gate allows to independently tune the gap and the position of the Fermi level in the regions underneath the gates 
 \cite{Overweg.2018}, such that a confining potential near the sample boundary with electrostatic control over its  slope is experimentally feasible.

\acknowledgements
Y.G. and B.R. acknowledge support by DFG RO 2247/11-1.   Y.G. was supported by CRC 183 (project C01), the Minerva Foun-
dation, MI 658/10-2, the
German Israeli Foundation (Grant No. I-118-303.1-2018),
the Helmholtz International Fellow Award, and by the
Italia-Israel QUANTRA grant.

\FloatBarrier
\bibliographystyle{apsrev4-2}

\ifarXiv
    

\section*{Supplementary material}

# Metal-insulator transition in a boundary three chain model
## Supplemental Material

Niels John,[1] Yuval Gefen,[2] and Bernd Rosenow[1]

[1]*Institut für Theoretische Physik, Universität Leipzig, D-04103, Leipzig, Germany*
[2]*Department of Condensed Matter Physics, Weizmann Institute, 7610001 Rehovot, Israel*

## NON-INTERACTING SYSTEM WITH $U = 0$

To diagonalize the Hamiltonian Eq. (1) with $U = 0$, we apply the Fourier transformation

$$c_{j,p,\sigma} = \frac{1}{\sqrt{N}} \sum_k e^{ijk} c_{k,p,\sigma} \; , \tag{S1}$$

and diagonalize the remaining Hamiltonian in $p$-space afterwards. The operators are transformed as

$$\begin{pmatrix} c_{k,+1,\sigma} \\ c_{k,0,\sigma} \\ c_{k,-1,\sigma} \end{pmatrix} = \begin{pmatrix} \sqrt{\frac{\tilde{V}^2+\sqrt{\tilde{V}^2+2}\tilde{V}+1}{2\tilde{V}^2+4}} & -\frac{\sqrt{\tilde{V}^2+2}+\tilde{V}}{\sqrt{2}\sqrt{(\tilde{V}^2+2)(\tilde{V}(\sqrt{\tilde{V}^2+2}+\tilde{V})+1)}} & \frac{1}{\sqrt{2}\sqrt{(\tilde{V}^2+2)(\tilde{V}(\sqrt{\tilde{V}^2+2}+\tilde{V})+1)}} \\ -\frac{1}{\sqrt{\tilde{V}^2+2}} & -\frac{\tilde{V}}{\sqrt{\tilde{V}^2+2}} & \frac{1}{\sqrt{\tilde{V}^2+2}} \\ \sqrt{\frac{\tilde{V}^2-\sqrt{\tilde{V}^2+2}\tilde{V}+1}{2\tilde{V}^2+4}} & \frac{\sqrt{\tilde{V}^2+2}-\tilde{V}}{\sqrt{2}\sqrt{(\tilde{V}^2+2)(\tilde{V}^2-\sqrt{\tilde{V}^2+2}\tilde{V}+1)}} & \frac{1}{\sqrt{2}\sqrt{(\tilde{V}^2+2)(\tilde{V}^2-\sqrt{\tilde{V}^2+2}\tilde{V}+1)}} \end{pmatrix} \begin{pmatrix} c_{k,1,\sigma} \\ c_{k,2,\sigma} \\ c_{k,3,\sigma} \end{pmatrix} , \tag{S2}$$

where $\tilde{V} = V/t$ is the scaled strength of the confining potential. The new bands have an index $\tau = -1, 0, +1$ and the respective Fermi momenta are $k_{F,\tau} = \arccos(-\mu/2t + \tau\sqrt{2t^2 + V^2}/2t)$. At half filling, $V = \sqrt{2}$ is an upper bound for the strength of the confining potential. Otherwise, the expression given before is ill defined. In physical terms, $V/t = \sqrt{2}$ marks the point where the upper band with $\tau = +1$ is empty and the lower band with $\tau = -1$ is completely filled. Consequently, only the middle band with $\tau = 0$ has Fermi points at $k_{F,0} = \pm\pi/2$. This is crucial for bosonization, since the starting point of the whole procedure are band dispersions, which we can linearize around their respective Fermi points [S1].

## MEAN FIELD ANALYSIS

We are interested in antiferromagnetic solutions and rewrite the local spin density as

$$n_{j,p,\sigma} = \frac{n_p}{2} - \sigma S_p (-1)^j \; , \tag{S3}$$

where $n_p$ is the average electron density and $S_p$ the staggered magnetization of chain $p$. Substitung Eq. (S3) into Eq. (1) and applying the standard mean field decoupling $n_{j,\uparrow} n_{j,\downarrow} \mapsto n_{j,\downarrow}\langle n_{j,\uparrow}\rangle_0 + n_{j,\uparrow}\langle n_{j,\downarrow}\rangle_0 - \langle n_{j,\uparrow}\rangle_0\langle n_{j,\downarrow}\rangle_0$, where the expectation value is respect to the Hartree-Fock ground state, we obtain the mean field Hamiltonian $H_{\text{MF}} = H_1 + H_2 + H_3 + H_\perp + H_C$ with

$$H_p = \sum_\sigma \sum_{k \in I} \epsilon_k \left( c^\dagger_{k,p,\sigma} c_{k,p,\sigma} - c^\dagger_{k-\pi,p,\sigma} c_{k-\pi,p,\sigma} \right) + U \frac{n_p}{2} \sum_\sigma \sum_{k \in I} \left( c^\dagger_{k,p,\sigma} c_{k,p,\sigma} + c^\dagger_{k-\pi,p,\sigma} c_{k-\pi,p,\sigma} \right)$$
$$+ U S_p \sum_\sigma \sum_{k \in I} \sigma \left( c^\dagger_{k\sigma} c_{k-\pi,p,\sigma} + c^\dagger_{k-\pi,p,\sigma} c_{k,p,\sigma} \right) - UN \left( \frac{n_p^2}{4} - S_p^2 \right) \tag{S4}$$

$$H_\perp = t \sum_{p=1,2} \sum_\sigma \sum_{k \in I} \left( c^\dagger_{k,p,\sigma} c_{k,p+1,\sigma} + c^\dagger_{k-\pi,p,\sigma} c_{k-\pi,p+1,\sigma} + \text{h.c.} \right) \tag{S5}$$

$$H_C = V \sum_\sigma \sum_{k \in I} \left( c^\dagger_{k,1,\sigma} c_{k,1,\sigma} + c^\dagger_{k-\pi,1,\sigma} c_{k-\pi,1,\sigma} - c^\dagger_{k,3,\sigma} c_{k,3,\sigma} - c^\dagger_{k-\pi,3,\sigma} c_{k-\pi,3,\sigma} \right) \tag{S6}$$

where $\epsilon_k = -2t\cos(k)$ and $I = [0, \pi]$. We denote the selfconsistent dispersions by $E_{k,\alpha,\gamma} \equiv \alpha \cdot \sqrt{E_{k,\gamma}^2}$ with $\alpha = \pm 1$ and $\gamma = -1, 0, +1$. The corresponding eigenspinors are $\Psi_{\alpha,\gamma}^{\tau,p}(k,\sigma)$. We note that the dispersion are doubly spin degenerate and $\alpha$ distinguishes between particle- and hole-like states. The observables $n_p$ and $S_p$ in terms of the dispersions and the eigenspinors are found as

$$n_p = \frac{1}{N} \sum_{k,\tau,\sigma} \sum_{\alpha,\gamma} \bar{\Psi}_{\alpha,\gamma}^{\tau,p}(k,\sigma) \Psi_{\alpha,\gamma}^{\tau,p}(k,\sigma) \cdot f_0(E_{k,\alpha,\gamma} - \mu) ,  \tag{S7}$$

and

$$S_p = -\frac{\sigma}{2N} \sum_{k,\tau} \sum_{\alpha,\gamma} \bar{\Psi}_{\alpha,\gamma}^{\tau,p}(k,\sigma) \Psi_{\alpha,\gamma}^{-\tau,p}(k,\sigma) \cdot f_0(E_{k,\alpha,\gamma} - \mu) . \tag{S8}$$

Here, $f_0(E)$ is the Fermi function and $\bar{\Psi}$ is the complex conjugate of $\Psi$. The chemical potential shows up explicitly in the equations,, whereas $V$ appears implicitly in the dispersions and the eigenspinors. The selfconsistent values of $n_p$ and $S_p$ are determined numerically for given $\mu$ and $V$. Figure 3 from the main text is obtained with the help of the selfconsistently determined dispersions.

## BOSONIZATION ANALYSIS

As outlined in the main text, we bosonize each band individually leaving us with the kinetic Hamiltonian

$$H_0 = \frac{1}{2\pi} \sum_\tau \sum_{\nu=\rho,\sigma} v_{F,\tau} \int dx \left[ (\partial_x \theta_{\nu,\tau})^2 + (\partial_x \phi_{\nu,\tau})^2 \right] , \tag{S9}$$

with $v_{F,\pm 1} = \sqrt{2 - (V/t)^2}$ and $v_{F,0} = 2$. We express the long wavelength density-density interaction terms as

$$H_{\text{int}}^0 = \frac{U}{\pi} \int dx\, (\partial_x \phi_\nu)^T \mathbf{H}_0 (\partial_x \phi_\nu) \tag{S10}$$

with $(\partial_x \phi_\nu) = (\partial_x \phi_{\nu,+1}, \partial_x \phi_{\nu,0}, \partial_x \phi_{\nu,-1})^T$ and

$$\mathbf{H}_0 = \begin{pmatrix} \frac{2\tilde{V}^4 + 4\tilde{V}^2 + 3}{2(\tilde{V}^2+2)^2} & \frac{2\tilde{V}^2+1}{(\tilde{V}^2+2)^2} & \frac{3}{2(\tilde{V}^2+2)^2} \\ \frac{2\tilde{V}^2+1}{(\tilde{V}^2+2)^2} & \frac{\tilde{V}^4+2}{(\tilde{V}^2+2)^2} & \frac{2\tilde{V}^2+1}{(\tilde{V}^2+2)^2} \\ \frac{3}{2(\tilde{V}^2+2)^2} & \frac{2\tilde{V}^2+1}{(\tilde{V}^2+2)^2} & \frac{2\tilde{V}^4+4\tilde{V}^2+3}{2(\tilde{V}^2+2)^2} \end{pmatrix} . \tag{S11}$$

Clearly, all of the three modes are coupled to each other in the Hamiltonian $\mathcal{H}_0 = H_0 + H_{\text{int}}^0$. Applying the transformation from Eq. (3) in the main text yields Eq. (4) with

$$\frac{u_\nu}{K_\nu} \hat{=} \begin{pmatrix} \sqrt{2-\tilde{V}^2} & 0 & 0 \\ 0 & 2 & 0 \\ 0 & 0 & \sqrt{2-\tilde{V}^2} \end{pmatrix} + \frac{U}{\pi} \begin{pmatrix} \frac{\tilde{V}^4+2\tilde{V}^2+3}{(\tilde{V}^2+2)^2} & \frac{\sqrt{2}(2\tilde{V}^2+1)}{(\tilde{V}^2+2)^2} & 0 \\ \frac{\sqrt{2}(2\tilde{V}^2+1)}{(\tilde{V}^2+2)^2} & \frac{\tilde{V}^4+2}{(\tilde{V}^2+2)^2} & 0 \\ 0 & 0 & \frac{\tilde{V}^2}{\tilde{V}^2+2} \end{pmatrix} . \tag{S12}$$

As outlined in the main text, we rescale the fields $\phi_{\nu,0}$ and $\theta_{\nu,0}$ such that $u_{\nu,\alpha} K_{\nu,\alpha} = \sqrt{2-(V/t)^2}\cdot$. Any orthogonal transformation will leave the matrix $u_\nu K_\nu$ invariant. Applying the final transformation from Eq. (9) yields the

diagonal Hamiltonian $\tilde{\mathcal{H}}_0$ from the main text with

$$\frac{\tilde{u}_{\nu,+}}{\tilde{K}_{\nu,+}} = \frac{U\tilde{V}^4 + 2U\sqrt{2-\tilde{V}^2}\tilde{V}^4 + 4U\sqrt{2-\tilde{V}^2}\tilde{V}^2 + 6U\sqrt{2-\tilde{V}^2} + 2U - 2\pi\left(\tilde{V}^2-3\right)\left(\tilde{V}^2+2\right)^2 - \sqrt{f(U,\tilde{V})}}{4\pi\sqrt{2-\tilde{V}^2}\left(\tilde{V}^2+2\right)^2}$$

$$\frac{\tilde{u}_{\nu,0}}{\tilde{K}_{\nu,0}} = \frac{U\tilde{V}^4 + 2U\sqrt{2-\tilde{V}^2}\tilde{V}^4 + 4U\sqrt{2-\tilde{V}^2}\tilde{V}^2 + 6U\sqrt{2-\tilde{V}^2} + 2U - 2\pi\left(\tilde{V}^2-3\right)\left(\tilde{V}^2+2\right)^2 + \sqrt{f(U,\tilde{V})}}{4\pi\sqrt{2-\tilde{V}^2}\left(\tilde{V}^2+2\right)^2} \quad (S13)$$

$$\frac{\tilde{u}_{\nu,-}}{\tilde{K}_{\nu,-}} = \sqrt{2-\tilde{V}^2} + \frac{U}{\pi}\frac{\tilde{V}^2}{\tilde{V}^2+2} \ .$$

with

$$\begin{aligned} f(U,\tilde{V}) = U^2 \Big[ &-4\tilde{V}^{10} - \left(4\sqrt{2-\tilde{V}^2}+7\right)\tilde{V}^8 - 8\left(\sqrt{2-\tilde{V}^2}+1\right)\tilde{V}^6 + 4\left(11\sqrt{2-\tilde{V}^2}+9\right)\tilde{V}^4 \\ &+ 12\left(4\sqrt{2-\tilde{V}^2}+5\right)\tilde{V}^2 - 8\sqrt{2-\tilde{V}^2}+76 \Big] \\ &- 4\pi U\left(\tilde{V}^2-1\right)\left(\left(2\sqrt{2-\tilde{V}^2}-1\right)\tilde{V}^4 + 4\sqrt{2-\tilde{V}^2}\tilde{V}^2 + 6\sqrt{2-\tilde{V}^2}-2\right)\left(\tilde{V}^2+2\right)^2 \\ &+ 4\pi^2\left(\tilde{V}^2-1\right)^2\left(\tilde{V}^2+2\right)^4 \ . \end{aligned} \quad (S14)$$

The individual Luttinger parameters and excitation velocities are calculated with the help of Eqs. (S13), (S14), and the relation $u_{\nu,\alpha}K_{\nu,\alpha} = \sqrt{2-(V/t)^2}$. as stated earlier in the text.

## UMKLAPP TERMS

As outlined in the main text, for the Umklapp terms the field chiralities of the field operators have to be chosen such that the displacement of the cosines is $\delta = \pm 2\pi$. We find four of such terms:

$$\begin{aligned} \mathcal{H}_{\text{int}}^{2k_F} = &-\frac{g_3^{(0,0;0,0)}}{2\pi^2\alpha^2}\cos\left(2\sqrt{2}\phi_{\rho,0}-\delta_0\cdot x\right) \\ &-\frac{g_3^{(1,-1;-1,1)}}{2\pi^2 a^2}\cos(2\theta_{\sigma,-})\cos(2\phi_{\rho,+}-\delta_+\cdot x) \\ &-\frac{g_3^{(1,-1;1,-1)}}{2\pi^2\alpha^2}\cos(2\theta_{\rho,-})\cos(2\phi_{\rho,+}-\delta_+\cdot x) \\ &+\frac{g_3^{(0,0;1,-1)}}{2\pi^2\alpha^2}\Big[\cos\left(\theta_{\rho,-}-\theta_{\sigma,-}\right)\cos\left(\phi_{\rho,+}-\phi_{\sigma,+}+\sqrt{2}\phi_{\rho,0}+\sqrt{2}\phi_{\sigma,0}\right) \\ &\qquad\qquad + \cos\left(\theta_{\rho,-}+\theta_{\sigma,-}\right)\cos\left(\phi_{\rho,+}+\phi_{\sigma,+}+\sqrt{2}\phi_{\rho,0}-\sqrt{2}\phi_{\sigma,0}\right)\Big] \end{aligned} \quad (S15)$$

| $g_3^{(0,0;0,0)}$ | $g_3^{(1,-1;1,-1)}$ | $g_3^{(1,-1;-1,1)}$ |
|---|---|---|
| $\frac{U}{t} \frac{(V/t)^4+2}{\left((V/t)^2+2\right)^2}$ | $\frac{U}{t} \frac{3}{2\left((V/t)^2+2\right)^2}$ | $\frac{U}{t} \frac{3}{2\left((V/t)^2+2\right)^2}$ |

TABLE S1. Bare values of the interaction constants. The lowest value of $U$ for which $g < 1/2$ is $U/t = 1$.

and the bare values of the interaction couplings are shown in Table S1. The scaling dimensions in first non-vanishing order is just the dimension of the operator itself. We obtain for the scaling dimensions

$$\frac{dg_3^{(0,0;0,0)}(l)}{dl} = \left(2 - 2(\eta_{\rho;0+})^2 \tilde{K}_{\rho,+} - 2(\eta_{\rho;00})^2 \tilde{K}_{\rho,0}\right) g_3^{(0,0;0,0)}(l)$$

$$\frac{dg_3^{(1,-1;-1,1)}(l)}{dl} = \left(2 - 1/\tilde{K}_{\sigma,-} - (\eta_{\rho;++})^2 \tilde{K}_{\rho,+} - (\eta_{\rho;+0})^2 \tilde{K}_{\rho,0}\right) g_3^{(1,1;-1,-1)}(l)$$

$$\frac{dg_3^{(1,-1;1,-1)}(l)}{dl} = \left(2 - 1/\tilde{K}_{\rho,-} - (\eta_{\rho;++})^2 \tilde{K}_{\rho,+} - (\eta_{\rho;+0})^2 \tilde{K}_{\rho,0}\right) g_3^{(1,-1;1,-1)}(l) \quad \text{(S16)}$$

$$\frac{dg_3^{(0,0;1,-1)}(l)}{dl} = \Bigg(8 - 1/\tilde{K}_{\rho,-} - 1/\tilde{K}_{\sigma,-} - (\eta_{\rho;++})^2 \tilde{K}_{\rho,+} - (\eta_{\rho;+0})^2 \tilde{K}_{\rho,0} - (t_\sigma^{(+,+)})^2 \tilde{K}_{\sigma,+} - (t_\sigma^{(+,0)})^2 \tilde{K}_{\sigma,0}$$

$$- \frac{(\eta_{\rho;0+})^2 \tilde{K}_{\rho,+} + (\eta_{\rho;00})^2 \tilde{K}_{\rho,0} + (t_\sigma^{(0,+)})^2 \tilde{K}_{\sigma,+} + (t_\sigma^{(0,0)})^2 \tilde{K}_{\sigma,0}}{\sqrt{2-(V/t)^2}}\Bigg) g_3^{(0,0;1,-1)}(l)/4$$

If we compare the respective scaling dimensions, we find that the $g_3^{(0,0;0,0)}$ and $g_3^{(1,-1;-1,1)}$ are more relevat than the other two sine-Gordon terms.

## COMPRESSIBILITY AND CONDUCTANCE

In the main text, we have used the fact that a gap in the bosonic excitation spectrum of a mode indicates that this mode is incompressible and does not contribute to the conductance. To make our manuscript self-contained, we describe this connection here. In Eq. (8), the charge density has been expressed in terms of the fields $\phi_{\rho,0}$ and $\phi_{\rho,+}$. In the case of RG relevant cosine terms Eq. (6), these fields are locked into an extremum of the cosine and acquire gaps according to Eq. (11). To be specific, we here focus on the field $\phi_{\rho,0}$, which is gapped for $V/t < 0.7$ and ungapped for larger $V/t$. We consider the following effective Hamiltonian for $\phi_{\rho,0}$

$$H = \int dx \left[\frac{1}{2\kappa}\rho_0^2(x) - \mu(x)\rho_0(x) + \Delta_0 \phi_{\rho,0}^2(x)\right] \quad , \tag{S17}$$

where $\rho_0 = \frac{\sqrt{2}}{\pi}\partial_x\phi_0$ and $\kappa^{-1} = \pi\hbar u/K$. In order to find the change in charge density due to an external chemical potential $\mu(x)$, we extremize the energy with respect to $\phi_{\rho_0}$, differentiate with respect to $x$, and Fourier transform to momentum space

$$\rho(q) = \frac{q^2 \kappa}{\kappa \pi^2 \Delta_0/2 + q^2} \mu(q) \quad . \tag{S18}$$

We see that in the limit $q \to 0$ there is no change in density when changing the chemical potential, hence the system is incompressible. In the absence of a gap however, $\rho(q) = \kappa\mu(q)$ as expected for a compressible system.

Following the derivation described in [S1], we obtain for the conductance of a system with size $L$

$$G = \frac{e^2}{\hbar\pi} \lim_{\omega \to 0} \frac{1}{\omega + i\delta} \int_{-\pi/a}^{\pi/a} \frac{dq}{2\pi} K(q,L) \left(\frac{\omega_n^2 uK}{\omega_n^2 + u^2 q^2 + \pi^2 \kappa \Delta_0/2}\right)_{i\omega_n \to \omega+i\delta} \tag{S19}$$

with $K(q,L) = \sin(qL/2)/(qL/2)$. Rescaling momenta according to $q \to q/\omega_n$, and performing the integral in the limit of small $\omega_n$, we finally obtain

$$G = -\frac{e^2}{\hbar\pi} \frac{K}{2\pi u} \frac{1}{\pi\sqrt{\kappa\Delta_0/2}} 2\arctan\left(\frac{1}{a\sqrt{\kappa\Delta_0/2}}\right) \lim_{\omega \to 0} -i(\omega + i\delta) = 0 \quad , \tag{S20}$$

i.e. the conductance vanishes. For $\Delta_0 = 0$ however, one obtains a finite conductance in the ususal way.

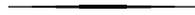


\fi

\end{document}